\documentclass[twocolumn,prl,aps,preprintnumbers,amsmath,amssymb,footinbib]{revtex4}
\usepackage{graphicx}% Include figure files
\usepackage{dcolumn}% Align table columns on decimal point
\usepackage{bm}% bold math
\usepackage{epsfig}
\usepackage{amssymb,amsmath,amsfonts,amsbsy,epsfig}%,psfrag,axodraw}
\usepackage{color}
\usepackage{times}

\def\lsim{\mathrel{\rlap{\lower4pt\hbox{\hskip1pt$\sim$}}
    \raise1pt\hbox{$<$}}}         %less than or approx. symbol
\def\gsim{\mathrel{\rlap{\lower4pt\hbox{\hskip1pt$\sim$}}
    \raise1pt\hbox{$>$}}}
\def\fun#1#2{\lower3.6pt\vbox{\baselineskip0pt\lineskip.9pt
  \ialign{$\mathsurround=0pt#1\hfil##\hfil$\crcr#2\crcr\sim\crcr}}}

\input epsf

\def\beq{\begin{equation}}
\def\eeq{\end{equation}}
\def\bea{\begin{eqnarray}}
\def\eea{\end{eqnarray}}

\begin{document}

\preprint{IPMU-10-0216}
%\preprint{UCB-xx-xxxx}
%\preprint{LBL-xx-xxxx}
\title{Saving fourth generation and baryon number by living long}
\author{Hitoshi Murayama~$^{\bf a,b,c}$}
% \email{murayama@berkeley.edu}
\author{Vikram Rentala~$^{\bf d, e}$}
% \email{rentala@berkeley.edu}
\author{Jing Shu~$^{\bf b}$}
% \email{jing.shu@ipmu.jp}
\author{Tsutomu T. Yanagida~$^{\bf b}$}
% \email{tsutomu.tyanagida@ipmu.jp}
\affiliation{
$^{\bf a}$ Department of Physics, University of California, Berkeley, CA 94720. \\
$^{\bf b}$ Institute for the Physics and Mathematics of the Universe,
University of Tokyo, Kashiwa, Japan  277-8568\\
$^{\bf c}$ Theoretical Physics Group, Lawrence Berkeley National Laboratory,
  Berkeley, CA 94720 \\
$^{\bf d}$ Department of Physics, University of Arizona, Tucson, AZ 85721 \\
$^{\bf c}$ Department of Physics, University of California, Irvine, CA 92697.
}

\begin{abstract}
  Recent studies of precision electroweak observables have led to the
  conclusion that a fourth generation %\blue{cannot exist}.
  is highly constrained.  However, we point out that a long-lived fourth
  generation can reopen a large portion of the parameter space.
%  is phenomenologically viable.
  In addition, it preserves baryon and lepton asymmetries against
  sphaleron erasure even if $B-L=0$.  It opens up the possibility of
  exact $B-L$ symmetry and hence Dirac neutrinos.  The fourth
  generation can be observed at the LHC with unique signatures of long-lived
  particles in the near future.
  \end{abstract}
\maketitle

When the muon was discovered as an exact copy of the electron but with a
higher mass, people wondered why nature repeats in an apparently
unnecessary fashion.  Later, discovery of CP violation led Kobayashi
and Maskawa to predict that nature actually repeats itself {\it at
  least}\/ three times.  There is no obvious reason why it should stop
with three.  At the same time, CP violation also led Sakharov to
consider how the apparent lack of anti-matter in the universe might be
explained. Therefore the apparent repetition of generations of
elementary particles has an intimate connection with the issue of
baryogenesis.

The fourth generation (4G) is indeed the simplest extension of the
standard model being searched for at Tevatron and at the LHC.  However,
several groups have claimed recently that this simple extension of the
standard model (SM) is highly constrained \cite{Erler:2010sk, Dawson:2010jx, Eberhardt:2010bm} or
already ruled out with no (CKM) mixing to the SM \cite{Chanowitz:2010bm}
by a combination of collider searches for its direct production and
its indirect effects in Higgs boson production, together with the
precision electroweak observables.

In this letter, we consider a long-lived 4G due to extremely small
mixings between the fourth and lighter three generations.  It could be
a consequence of a flavor symmetry or compositeness of the 4G.  We
then point out that such a 4G evades these constraints.  This is because
the 4G neutrino can be nearly stable, and it can be below the nominal
LEP-II limit. Its loops can generate negative $S$, which in turn
allows for a heavier Higgs boson, and positive $T$ for a small
splitting between 4G up- and down-type quarks.

Interestingly, such a long-lived 4G has an important implication for
baryogenesis.  The baryon number $B$ is usually believed to be erased
unless there is a non-vanishing asymmetry in $B-L$.  However,
the longevity of the 4G provides additional conserved numbers beyond
$B-L$, which in turn protects $B$ from erasure. Therefore, the
longevity leads to both 4G asymmetry and $B$ asymmetry. It even allows for an
exact $B-L$ symmetry of nature, either global or local, making us reconsider
the origin of neutrino mass and baryon asymmetry.

Let us first discuss the current constraints on the 4G particles, with
the obvious notation $U$, $D$, $E$, and $N$.  The electroweak
precision tests (EWPTs) for the chiral 4G are summarized in
Ref.~\cite{Kribs:2007nz} with some more recent updates in
Ref.~\cite{Erler:2010sk, Eberhardt:2010bm}. Here we use the latest
global fit results from Ref.~\cite{Erler:2010sk} which include the
constraints from the low-energy data. The $S$-$T$ ellipse in ~\cite{Erler:2010sk} looks
somewhat more horizontal (large positive $T$ is not preferred now)
than the one used in
Ref.~\cite{Kribs:2007nz} from the LEP Electroweak Working Group \cite{EWWG}. As a consequence, some of the sample points
in Ref.~\cite{Kribs:2007nz} do not lie in the 95\%~C.L. in the
$S$-$T$ plot and the allowed parameter space for $m_{U}, m_{D}$ is
much smaller.

If the 4G Dirac neutrino decays, it has to be heavier than 90.3~GeV
\cite{PDG08}.  However if it is long lived at LEP, then the only bound
is $m_{N} > 45.0$~GeV from the invisible $Z$ decay width.  For the
charged leptons, the bound is $m_{E} > 102.6$ GeV for the long-lived
case and $m_{E} > 100.8$ GeV for the shorted-lived case
\cite{PDG08}. Unlike all the previous papers, which assumed $m_{E,N}
\gtrsim 100$~GeV, we scan over all possible 4G lepton masses. All the
$S$, $T$ contributions from 4G fermions are calculated from the exact
one-loop formulae in Ref.~\cite{He:2001tp} while the two-loop Higgs
contribution is obtained from fitting the previous results. The Higgs
mass is chosen within the two allowed mass regions, $m_{h} = 130 <
131$~GeV (light) and $m_{h} = 300 > 204$~GeV (heavy) at 95\%~C.L. from
the latest Higgs boson search at the Tevatron~\cite{Aaltonen:2010sv}
that includes the loop of the 4G in the gluon-fusion process.

There are also lower bounds from direct searches for the
fourth-generation quarks at the Tevatron. In the long-lived case, we
infer the bounds from the limit on stable stop
\cite{Aaltonen:2009kea}. Rescaling the production cross section with
the $U \bar{U}$ production rate at NLO level \cite{tp} gives us $m_{D}
\geqslant m_{U} > 340$ GeV and $m_{U} \geqslant m_{D} > 310$ GeV at
95\% C. L. For the short-lived case, we obtain $m_{U} > 335$ GeV from
$W +$ jets \cite{tp} if $m_{D} \geqslant m_{U}$ and $m_{D} > 338$ GeV
if $m_{U} \geqslant m_{D}$ and $D$ decays dominantly into $W + t$
\cite{Aaltonen:2009nr}.  The bound does not change significantly for a
sizable branching ratio $D \rightarrow W + $ jets
\cite{Flacco:2010rg}.

There are also upper bounds for the $Q=U$ or $D$ mass from tree
level unitarity. The most stringent bounds are from the scattering $Q
\bar{Q} \rightarrow Q \bar{Q} $ which includes the color off-diagonal
amplitudes \cite{Chanowitz:1978mv}. Requiring the eigenvalue of the
tree-level partial-wave amplitude to be smaller than 1/2, we find
\begin{equation}
\frac{2 \sqrt{2} \pi }{G_F} > [ 3 (m_U^2 + m_D^2 ) + \sqrt{9 (m_U^2 -
  m_D^2)^2 + 16 m_U^2 m_D^2 } ] \ .
\end{equation}
Regarding the unitarity limit, remember that the amplitude is only
calculated at the tree level and hence the bound is soft.

\begin{figure}
%  \includegraphics[width=8cm]{mtmb1}
  \includegraphics[width=8cm]{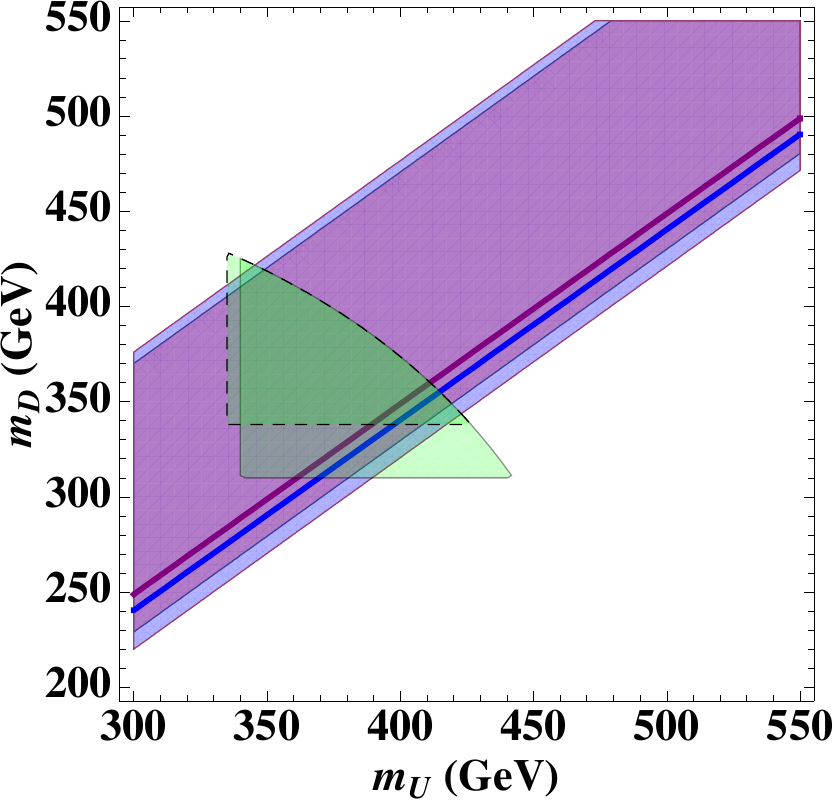}
  \caption{The $m_{D}$ vs $m_{U}$ contour plot for varying
    fourth-generation lepton masses. The purple region is the allowed
    mass region from the $S$-$T$ constraint at 95\% C.L. for $m_h =
    130$ GeV and the blue region (including the purple region) is that
    for $m_h = 300$ GeV. The lower limit of the green region comes
    from the direct searches at the Tevatron (The solid line is the
    case for long-lived fourth-generation quarks and the dashed line
    is the case for the prompt decay), while the upper limit is the
    bound from unitarity. The purple and the blue lines 
    use the approximate formula for the fourth generation quark masses
    $m_{U} - m_{D} = (1+ \frac{1}{5} \log(\frac{m_h}{115 GeV})) *50$
    GeV from Ref.~\cite{Kribs:2007nz} for $m_h = 130$ GeV and 300 GeV,
    respectively.}
  \label{fig:mtmb}
\end{figure}

The allowed mass region for the fourth-generation quarks is presented
in Fig.~\ref{fig:mtmb}. Firstly and most importantly, we find a large
allowed region for $m_{U} \lesssim m_{D}$ as opposed to previous
analyses which assumed $m_{E,N} \gtrsim 100$~GeV, whose contribution
to the $S$ parameter is positive together with the Higgs
contribution. Then the $S$-parameter constraint only allows a small
mass region for $m_{U} > m_{D}$. However, for a light
fourth-generation neutrino (around 46 GeV), the fourth-generation
lepton contribution to the $S$ parameter is negative (around
$-0.09$). Hence, large fourth-generation quark masses ($m_{U}
\lesssim m_{D}$) with a relatively large $S$ parameter are
allowed. Secondly, varying $m_h$ does \textsl{not}\/ change the
allowed parameter space because decreasing $m_{N}$ can compensate for
the $S$, $T$ contribution from increasing the Higgs mass.

%Here we choose two sample fourth generation lepton masses. One is $m_{N} = 100$~GeV, $m_{E} = 155$~GeV which is chosen in Ref. \cite{Kribs:2007nz}. The other one is the extremely light case $m_{N} = 51$~GeV, $m_{E} =103$~GeV (notice their contribution to $S$ is negative which allows more room for the mass of fourth generation quarks).

In Ref.~\cite{Chanowitz:2010bm}, it is claimed that the 4G with small
mixing is ruled out by the a combination of EWPTs, direct searches and
the indirect bounds from the Higgs production at Tevatron. We will explain our disagreement with the author.
The author first fixes $m_{U}-m_{D} = 16$ GeV and $m_{E} - m_{N}$ = 91 GeV
\cite{Erler:2010sk} which is a very limited region of the allowed
parameter space which is clear from our Fig.~\ref{fig:mtmb}. Then he
finds the Higgs mass required by EWPTs in the zero mixing case for the
two end point mass of $m_{U}$ and $m_{D}$ in the line $m_{U}-m_{D} =
16$ GeV is ruled out by the direct Tevatron Higgs boson search.
However, the mixing between 3rd and 4th generation increases the $T$
parameter only (see formula one in Ref.~\cite{Chanowitz:2010bm}), effectively the same as
increasing the ${U}$ and ${D}$ mass splitting \endnote{The Yukawa
  couplings hit the Landau pole below 100~TeV or so.  We will not
  discuss its UV completion above this energy scale as it is
  irrelevant to the following discussions.}.

Having demonstrated that longevity makes the 4G
phenomenologically viable, we now turn our attention to the baryon
asymmetry. In the SM, $B$ and $L$ are separately conserved except for the sphaleron transitions \cite{Klinkhamer:1984di} which violate $B+L$ but preserve $B-L$. The net $B$ must be proportional to the only conserved quantity $B-L$ and would be zero if $B-L=0$. After the the sphaleron transitions get decoupled from thermal equilibrium at $T_\textrm{sph}$, $B$ is a conserved quantity which gives us the right number density today.  

If the 4G fermions do not mix with the lighter generations significantly, they would stay in chemical equilibrium only through the electroweak sphaleron transitions, which maximally violate $3B+3L + B_4 + L_4$ instead. Here we use $B$ or $L$ as the baryon or lepton number in the first three generations. In this case, the other three orthogonal combinations of $B$, $L$, $B_4$, $L_4$ are conserved charges. The final $B$ is a linear combination of these conserved charges instead of just being proportional to $B-L$. As a consequence, unless there exists some accidental cancellations, the net baryon number density would be nonzero even if $B-L = 0$.  

The thermal history for baryon number generation after inflation is
summarized as follows. First, we assume there is some baryogenesis
mechanism which generates a net baryon asymmetry $B+B_4 =L +L_4\neq 0$. However, this initial condition could generate an asymmetry in the other conserved charges (for instance, $L - 3 L_4$). Above the critical temperature $T_c$ of the electroweak phase transition, all particles are massless and net $B$ could be small or zero depending on the particle content of the model.
Below $T_c$, all fermions gain their masses via the Higgs mechanism, so it costs additional energy to create the heavy fourth generation fermions. Once the temperature drops below their masses, the mass effect essentially blocks the sphaleron process from erasing $B$ \footnote{There are related scenarios which generates the
      dark matter abundance through sphalerons \cite{Buckley:2010ui},
      preserves $B$ relying on $\tau$ lepton mass \cite{Shu:2006mm} or
      Dirac neutrino mass \cite{Dick:1999je}.}.

We follow the standard analysis in Ref.~\cite{Harvey:1990qw} while
taking into account all the mass
effects. %The accumulated evidence for the lepton mixings suggest that
We choose a single chemical potential $\mu_l$ for leptons instead of
separate chemical potentials for each light lepton flavor as in
Refs.~\cite{Kuzmin:1987wn, Dreiner:1992vm, Shu:2006mm,
  Carroll:2005dj}. We consider the SM matter consisting of three
families, each of which consists of two quarks (an up-type and
down-type) with masses $m_{q_i}$, a charged lepton of mass $m_{l_i}$
and a massless
neutrino. %which for our purposes can be approximated as massless.
The SM interactions relate all the chemical potentials which leave us with
six independent chemical potentials in our case: $\mu_{u_L}$, $\mu_W$,
$\mu_0$, $\mu_{U_L}$, $\mu_{N_L}$, $\mu = \sum_i \mu_i =3 \mu_{\nu_L}$
which are the chemical potentials for upper type quarks, $W^-$ bosons,
neutral Higgs boson, 4G up type quark, 4G neutrino, sum over all SM
neutrino chemical potentials.
\begin{eqnarray}
% \mu_W &=& \mu_{-} + \mu_0  ~~~~~~  (W^{-} \leftrightarrow \phi^- + \phi^0)  \nonumber \\
\mu_{d_L} &=& \mu_{u_L} + \mu_W ~~~ (W^{-} \leftrightarrow \bar{u}_L + d_L)  \nonumber \\
\mu_{D_L} &=& \mu_{U_L} + \mu_W ~~~ (W^{-} \leftrightarrow \bar{U}_{L} + D_{L})  \nonumber \\
\mu_{e_L} &=& \mu_{\nu_L} + \mu_W ~~~~~ (W^{-} \leftrightarrow \bar{\nu}_{L} + e_{L} )  \nonumber \\
\mu_{E_L} &=& \mu_{N_L} + \mu_W ~~~~~ (W^{-} \leftrightarrow \bar{N}_{L} + E_{L} )  \nonumber 
\\
%\mu_{iL} &=& \mu_{i} + \mu_W ~~~~~ (W^{-} \leftrightarrow \bar{\nu}_{i L} + e_{i L} )  \nonumber 
\mu_{u_R} &=& \mu_{0} + \mu_{u_L} ~~~~~~~ (\phi^0 \leftrightarrow \bar{u}_L + u_R)  \nonumber 
\end{eqnarray}
\begin{eqnarray}
\mu_{U_R} &=& \mu_{0} + \mu_{U_L} ~~~~~~~ (\phi^0 \leftrightarrow \bar{U}_{L} + U_{R})  \nonumber 
\\
\mu_{d_R} &=& - \mu_{0} + \mu_W + \mu_{u_L} ~~~ (\phi^0 \leftrightarrow \bar{d}_L + \bar{d}_R)  \nonumber \\
\mu_{D_R} &=& - \mu_{0} + \mu_W + \mu_{U_L} ~~~ (\phi^0 \leftrightarrow \bar{D}_{L} + \bar{D}_{R})  \nonumber \\
\mu_{e_R} &=& - \mu_{0} + \mu_W + \mu_i ~~~~~ (\phi^{0} \leftrightarrow e_{L} + \bar{e}_{R})
 \nonumber \\
\mu_{E_R} &=& - \mu_{0} + \mu_W + \mu_{N_L} ~~~~~ (\phi^{0} \leftrightarrow E_{L} + \bar{E}_{R})
 \nonumber \\
\mu_{N_R} &=& - \mu_{0} + \mu_W + \mu_{N_L} ~~~~~ (\phi^{0} \leftrightarrow N_L + \bar{N}_{R})
\end{eqnarray}

% Those six independent chemical potentials we used in the paper are $\mu_{uL}$, $\mu_W$, $\mu_0$, $\mu_{u4_L}$, $\mu_{ \nu 4_L}$, $\mu = \sum_i \mu_i$ \blue{(The $\mu$ is 3 $\mu_{\nu L}$ here.)} which are the chemical potentials for upper type quarks, $W^-$ bosons, neutral Higgs bosons, 4th generation upper type quark, 4th generation neutrino, sum over all neutrino chemical potentials (we assume they are all equal here).

The mass correction to the particle number asymmetry density $n_p$ is
\begin{eqnarray}
\label{eq:massfun}
n_p &=& \frac{g_p}{\pi^2} T^3 \left( \frac{\mu}{T} \right ) \int^\infty_x y \sqrt{y^2 - x^2} \frac{e^y}{(1 \pm e^y)^2} d y \nonumber \\
&=& \left\{
 \begin{array}{l}
  \displaystyle \frac{g_p T^3}{3} \left( \frac{\mu}{T} \right ) \alpha_b(x)  ~~~
  \textrm{$p$ is a boson,}
  \\
  \displaystyle \frac{g_p T^3}{6} \left( \frac{\mu}{T} \right ) \alpha_f(x)  ~~~
  \textrm{$p$ is a fermion,}
  \end{array}\right.
\end{eqnarray}
where we assume $n_p \propto \mu$ for small asymmetries. $g_p$ is the number of internal degrees of freedom and $x =
m/T$. The mass correction functions for bosons and fermions are
normalized as $\alpha_b (0) = \alpha_f (0) = 1$. We define $\Delta
\equiv N - \sum_i \alpha_i$ $(N=3)$ for SM particles with $i=1, 2, 3$
generations. $\Delta_u$, $\Delta_d$ and $\Delta_i$ stands for the
overall mass corrections for up type SM quarks, down type SM quarks
and SM charged leptons,
respectively. %when summing over $N=3$ light generations.
%In the massless limit, $\Delta_u = \Delta_d = \Delta_i = 0$, $\alpha_W = \alpha_- = \alpha_0 = 1$.
The $\alpha_W$,
% $\alpha_-$,
$\alpha_0$, $\alpha_U$, $\alpha_D$, $\alpha_E$ and $\alpha_N$ are the
mass function in Eq. (\ref{eq:massfun}) for $W$ boson,
% charged Higgs,
neutral Higgs, 4G up-quark, 4G-down quark, 4G charged lepton and 4G neutrino respectively. It is easy to see $\Delta_d$ and $\Delta_i < 5 \times
10^{-4}$ since $T_{\textrm{sph}} > m_W$ so we will ignore their contribution in
the following discussions. The neutral Higgs boson condenses so we
have $\mu_0 = 0$. One can write the charge densities in terms of the chemical potential (upto irrelevant constants):
\begin{eqnarray}
% Q &\approx& 2 (N - 2 \Delta_u ) \mu_{u_L} - 2 ( 2 N + 2 \alpha_W + n \alpha_- ) \mu_W - 2 \mu \nonumber \\
Q &\approx& 2 (N - 2 \Delta_u ) \mu_{u_L} - 2 ( 2 N + 3 \alpha_W ) \mu_W - 2 \mu \nonumber \\
& & + 4 \alpha_{U} \mu_{U_L} - 2 \alpha_{D} (\mu_{U_L} + \mu_W) -2 \alpha_{E} (\mu_{N_L} + \mu_W) \nonumber \\
B &\approx& ( 4 N - 2 \Delta_u ) \mu_{u_L} +  2 N \mu_W
\nonumber \\
L &\approx& 3 \mu + 2 N \mu_W \nonumber \\
B_4 &=& 2 \alpha_{U} \mu_{U_L} + 2 \alpha_{D} (\mu_{U_L} + \mu_W)  \nonumber \\
L_4 &=& 2 \alpha_{N} \mu_{N_L} + 2 \alpha_{E} (\mu_{N_L} + \mu_W)  \ ,
\end{eqnarray}
where the net $Q$ (electric charge density) must be 0. The conserved charge densities are
\begin{eqnarray}
\label{eq:const23}
 B - L &=& ( 4 N - 2 \Delta_u ) \mu_{u_L} - 3 \mu  \nonumber \\
 B_4 - L_4 &=& 2 \alpha_{U} \mu_{U_L} + 2 \alpha_{D} (\mu_{U_L} + \mu_W)
\nonumber \\
& & - 2 \alpha_{N} \mu_{N_L} - 2 \alpha_{E} (\mu_{N_L} + \mu_W) \nonumber \\
 L - 3 L_4 &=& 3 \mu + 2 N \mu_W - 6 \alpha_{N} \mu_{N_L}  \nonumber \\
& & - 6 \alpha_{E} (\mu_{N_L} + \mu_W)  \ .
\end{eqnarray}
The electroweak sphaleron process which converts $qqql$ of each generation into nothing give us
the last constraint
\begin{eqnarray}
\label{eq:const1}
3 N \mu_{u_L} + 2 (N +1) \mu _W + \mu +3 \mu_{U_L} + \mu _{N_L} = 0 \ .
\end{eqnarray}

%%%%%%%%%%%%%%%
\begin{figure}[bt]%[htbp]
\centering
\includegraphics[width=0.45 \textwidth]{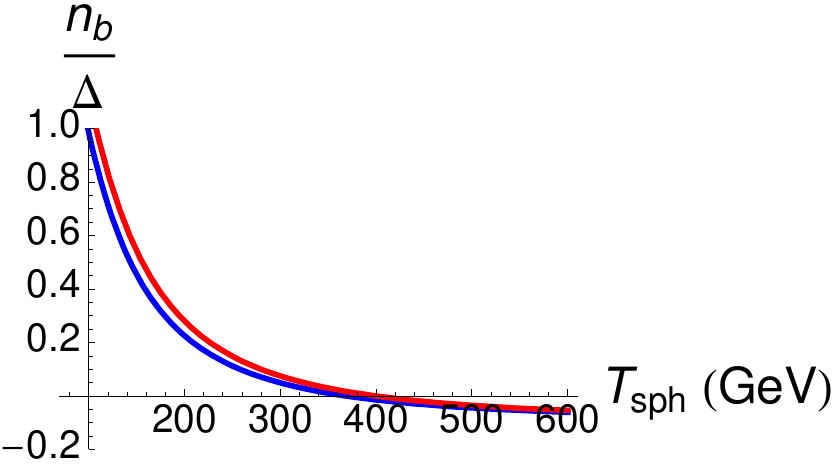}
\caption{\label{fig:ratioB1L1} The final baryon asymmetry versus the
  inital asymmetry $n_b / \Delta$ as as a function of sphaleron
  freeze-out temperature $T_{\textrm{sph}}$ (GeV). The blue (red) lines are for
  $m_{N} = 46 (46)$ GeV, $m_{E} = 134 (103)$ GeV, $m_{U} = 350 (380)$
  GeV, $m_{D} = 350 (380)$ GeV, $m_{\phi_0} = 300 (130)$ GeV. }
\end{figure}
%%%%%%%%%%%%%%

%\begin{eqnarray}
%\label{eq:massfun}
%\alpha_b(x) &=& \frac{3}{\pi^2} \int_x^\infty y \sqrt{y^2 - x^2} \frac{e^y}{(1-e^y)^2} dy \ , \nonumber \\
%\alpha_f(x) &=& \frac{6}{\pi^2} \int_x^\infty y \sqrt{y^2 - x^2} \frac{e^y}{(1+e^y)^2} dy \ ,
%\end{eqnarray}
%where we have normalized the functions so that $\alpha_b (0) = \alpha_f (0) = 1$. %The numerical results are presented in FIG. \ref{Fig:masscorrection}.

%%%%%%%%%%%%%%%
%\begin{figure}[bt]%[htbp]
%\centering
%\includegraphics[width=0.53 \textwidth]{masscorrection.pdf} \\
%\includegraphics[width=0.45 \textwidth]{masscorrection1}
%\caption{\label{Fig:masscorrection} The explicit mass corrections for
%  different spieces as a function of temperature $T$. The black, red,
%  green, blue lines are $\Delta_u$, $\alpha_W$, $\alpha_-$, $\alpha_0$
%  respectively. The top, $W$ boson, charged Higgs, neutral Higgs mass
%  are chosen to be 172, 80, 120, 130 GeV respectively. }
%\end{figure}
%%%%%%%%%%%%%%

%Since we are only interested in the final baryon number density nowadays, which is after  the electroweak symmetry breaking. The sphaleron effect we consider is in within $T_{sph} < T < T_c$, where $T_{sph}$ is the lower frozen out temperature for spharelon and $T_c$ is critical temperature for the electroweak phase transition (EWPT). In this case, we have $\mu_0 = 0$ due to the Higgs condensate.

Now we can pick up two sample spectra which are consistent with the
most recent data, $m_{N} = 46$ GeV, $m_{E} = 103$ GeV, $m_{U} = 380$
GeV, $m_{D} = 380$ GeV, $m_{\phi_0}$ = 130 GeV or $m_{N} = 46$ GeV,
$m_{E} = 134$ GeV, $m_{U} = 350$ GeV, $m_{D} = 359$ GeV, $m_{\phi_0}$
= 300 GeV and show how the final baryon asymmetry is obtained from an
initial baryon asymmetry with $B - L =0$. The full numerical results
including all the mass effects are presented in
FIG. \ref{fig:ratioB1L1}. We choose the initial asymmetry as $B$ = $L$
= 3 $\Delta$, $B_4$ = $L_4$ = 0 and use the minimal 4G, %$n=1$,
$m_t =$172 GeV, $m_W$ = 80 GeV. One can clearly see that the final
baryon number density is the same order as the initial baryon number
density 3 $\Delta$ if the sphaleron decoupling temperature is not too
high. Note that even for very high sphaleron decoupling temperature $T_\textrm{sph}$ when the 4G fermions are essentially massless, the baryon number is not completely erased because of the mismatch in the number of the neutrino degrees of freedom.

%\begin{figure}
%  \includegraphics[width=0.45 \textwidth]{survivalfraction}\\
%  \caption{Survival Fraction}\label{SF}
%\end{figure}

% \section{Experimental constraints.}
% \label{sec: constraints}

% \subsubsection{Electroweak Precision Tests}

% \section{LHC signals}
% \label{sec:LHC}

\begin{figure}
%  \includegraphics[width=8cm]{mtmb1}
  \includegraphics[width=8cm]{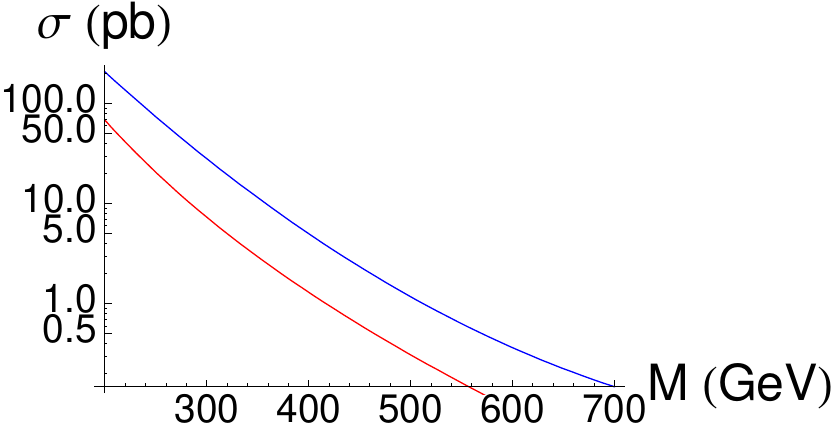}
  \caption{Production rate for the $U \bar{U}$ at the LHC, the blue
    curve is the cross section $\sigma(p p \rightarrow U \bar{U})$
    computed by PYTHIA at the LHC with $\sqrt{s} = 14$ TeV from
    Ref. \cite{Quertenmont:2008zza}. The red curve is the one at the
    early LHC with $\sqrt{s} = 7$ TeV at the NLO level from
    \cite{Berger:2009qy, qinghong}.}
  \label{fig:prod}
\end{figure}

For the LHC signals of the long-lived 4G quarks \footnote{The
  estimation of the proper lifetime and traveling distance also
  applies to the forth generation quarks, and the lightest fourth
  generation neutrino will look like missing energy.}, we first
estimate their proper lifetime. $U / D$ have to decouple from the SM
fermions above the sphaleron freeze-out temperature $T_{\textrm{sph}}$
which gives us the lower limit on the proper lifetime: $\Gamma (U /D
\rightarrow q W^{\pm} )< H\sim T_{\textrm{sph}}^2 / M_{\textrm{pl}}$. The
long-lived particles $U / D$ should not disrupt the success of
BBN which gives us the upper limit of the proper lifetime. Then the
allowed window for the proper lifetime is $10^{-10} \textrm{s} <
\tau_{Q} < 1 \textrm{s} $, which also corresponds to the small mixing
angle $10^{-13}< \theta < 10^{-8}$.  Their decay length at the LHC is
%\begin{eqnarray}
$d = \beta c \tau \gamma \approx (30 mm)
\left( \tau / 10^{-10} s\right) \beta \gamma $.
%\end{eqnarray}
%This opens up the possibility to give a strong hint for our scenario by detecting those long-lived charged particles.
If the lifetime is relatively short within the above range,
  the 4G quarks show displaced vertices in their decays.  On the other
hand, if the 4G quarks decay outside the detector,
%In this case,
the lighter 4G quark would hadronize and the
signal would look like a jet with tracks, with anomalously large energy
deposits in the silicon detector or delayed hits in the calorimeters
or muon chamber. At the early LHC, this is one of the signals that can
be looked for.  At the same time, it may cause confusion if the
charge-exchange reaction with the detector material causes the charged
bound state to turn neutral and vice versa, making the track a
``dashed line'' \cite{Fairbairn:2006gg}.

Unfortunately, we are not aware of any ATLAS/CMS simulation on the
long-lived 4G quarks. However, we can rescale the production rate and
use the study for the long-lived stop since the spin effect is negligible
for mesons or baryons with a heavy constituent. The production rate for
stop at the 14 TeV LHC and 4G $U$ at $\sqrt{s}=7$~TeV is presented in
Fig.~\ref{fig:prod}. We can see that for a typical mass range (300
$\sim$ 400 GeV) allowed by EWPTs, direct searches and unitarity limit,
the 4G $U$ production rate at $\sqrt{s} = 7$ TeV LHC is roughly 1/8th of
stop production rate at $\sqrt{s} = 14$ TeV LHC. In
Table~\ref{table1}, we list the several required integrated luminosity
$\mathcal{L}_{\textrm{int}}$ for LHC at $\sqrt{s} = 14$ TeV to observe 3 events
in CMS \cite{Rizzi:2010zz} for the long-lived stop production of
different masses. If we assume that the acceptance for the signals to
pass the cuts and the background are similar in the above two
situations, we can estimate the order of magnitude for the required
integrated luminosities to observe hints ({\it e.g.}\/, about three
events) at the LHC ($\sqrt{s} = 7$ TeV) is $\mathcal{L}_{\textrm{int}} = $ 10
$\sim$ 10$^2$ pb$^{-1}$ which is promising for the end of this
year. With more data accumulated at the $\mathcal{L}_{\textrm{int}} \simeq $ 1
fb$^{-1}$, we expected that there would be decisive evidence for such
unique signatures of long-lived particles.

\begin{table}[b]
  \caption{\label{table1}The required integrated luminosity
    $\mathcal{L}_{\textrm{int}}$ for LHC $\sqrt{s} = 14$ TeV to observe 3
    events for the long-lived stop production for their different
    masses. The data are quoted from Fig. 2 (left) in
    Ref. \cite{Rizzi:2010zz}.  }
\begin{ruledtabular}
\begin{tabular}[b]{c|c|c|c|c|c|c}
$\mathcal{L}_{\textrm{int}}$ (pb$^{-1}$) & 0.2 &  1  & 4 & 20 & 40 & 100 \\
\hline
$m$ (GeV) & 200   & 300 & 400 & 500 & 600 & 700
\end{tabular}
\end{ruledtabular}
\end{table}

%From Fig. \ref{fig:prod}, we can see that the roughly speaking, for the allowed masses for the fourth generation quarks, we can expect $10^2 \sim 10^3$ of them would be produced at the end of this year for $30 \sim 50$ pb$^{-1}$ integrated luminosity. Now the questions are the following parameters which have large uncertainties.

We would like to thank Qinghong Cao for providing the production rate
at the NLO level for light 4th generation quark at the early LHC. The
work is partially supported by the World Premier International
Research Center Initiative (WPI initiative) MEXT, Japan. H.M. was also
supported in part by the U.S. DOE under Contract DE-AC03-76SF00098, in
part by the NSF under grant PHY-04-57315, and in part by the
Grant-in-Aid for scientific research (C) 20540257 from Japan Society
for Promotion of Science (JSPS). J.S. is also supported by the
Grant-in-Aid for scientific research (Young Scientists (B) 21740169)
from JSPS.

\end{document}